\documentclass[aps,prb,twocolumn,amsmath,amssymb,eqsecnum,nofootinbib,superscriptaddress,floatfix,showpacs]{revtex4}

\usepackage{bm} 
\usepackage{graphicx}
\usepackage{amssymb}
\usepackage{curves}
\usepackage{amsmath}
\usepackage{fancybox}
\usepackage{dcolumn} 
\usepackage[dvips]{color}
\usepackage{epic}
\usepackage{longtable}

\begin{document}

\title{
Two-dimensional spin-filtered chiral network model for the 
$\mathbb{Z}^{\ }_2$ quantum spin-Hall effect
      }

\author{Hideaki Obuse}
\affiliation{Condensed Matter Theory Laboratory,
             RIKEN, 
             Wako, 
             Saitama 351-0198, 
             Japan}

\author{Akira Furusaki}
\affiliation{Condensed Matter Theory Laboratory,
             RIKEN, 
             Wako, 
             Saitama 351-0198, 
             Japan}

\author{Shinsei Ryu}
\affiliation{Kavli Institute for Theoretical Physics,
 	     University of California, 
 	     Santa Barbara, 
 	     California 93106, 
 	     USA}
\author{Christopher Mudry}
\affiliation{Condensed Matter Theory Group,
              Paul Scherrer Institute,
              CH-5232 Villigen PSI,
 	      Switzerland}

\date{\today}

\begin{abstract}
The effects of static disorder on the 
$\mathbb{Z}^{\ }_2$ quantum spin-Hall effect
for non-interacting electrons propagating in two-dimensional space
are studied numerically. A two-dimensional time-reversal symmetric 
network model is constructed to account for the effects of static disorder 
on the propagation of non-interacting electrons subjected to 
spin-orbit couplings. This network model is different from past network models 
belonging to the symplectic symmetry class in that 
the propagating modes along the links of the network can be
arranged into an odd number of Kramers doublet. It is found that 
(1) a two-dimensional metallic phase of finite extent 
is embedded in a $\mathbb{Z}^{\ }_{2}$ insulating phase in parameter space
and (2) the quantum phase transitions between the metallic and
$\mathbb{Z}^{\ }_{2}$ insulating phases belong to the conventional 
symplectic universality class in two space dimensions.
\end{abstract}


\maketitle

\section{
Introduction
        } 

An early triumph of quantum mechanics applied to the theory of solids
was the understanding that, in the thermodynamic limit,
the metallic state can be distinguished
from the insulating state based on the energy spectrum of non-interacting 
electrons subject to the (static) periodic crystalline potential.
The Bloch insulating state occurs when the chemical potential
falls within the energy gap between the electronic Bloch bands
while the metallic state follows otherwise.

It took another 50 years with the experimental discovery of the 
integer quantum Hall effect\cite{ReviewIQHE}  
to realize that a more refined
classification of the Bloch insulating state follows from the
sensitivity of occupied Bloch states to changes in the boundary conditions.
A two-dimensional electron gas subjected
to a strong magnetic field turns into a quantum Hall insulating state 
characterized by a quantized Hall conductance in units of $e^2/h$.%
\cite{Laughlin81,Halperin82,Thouless82,Avron83,Kohmoto85,Niu85,Arovas88,Hatsugai93} 
The topological texture of the quantum Hall insulating state 
manifests itself through the existence of chiral edge states:%
\cite{Halperin82}
energy eigenstates that propagate in one direction along
the boundary of a sample with strip geometry.
On the other hand, the topologically trivial Bloch 
insulating state is insensitive to 
modification of boundary conditions and, therefore,
it does not support gapless edge states in a strip geometry.
The breaking of time-reversal symmetry
by the magnetic field in the integer quantum Hall effect
implies the chirality of edge states: 
all edge states propagate 
in the same direction. 
Chiral edge states thus cannot be 
back-scattered into counter propagating edge states by impurities.
For this reason the quantization of the Hall conductance is insensitive
to the presence of (weak) disorder.%
\cite{Halperin82} (Strong disorder destroys the very existence of edge states.)

\begin{figure}[b]
  \begin{center}
  \includegraphics[width=0.2\textwidth]{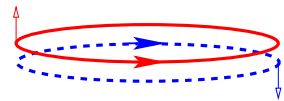}
\caption{
\label{fig: qshe}
(Color online:)
In the proposal of Ref.\ \onlinecite{Bernevig05},
a spin-up edge state (full line)
and a spin-down edge state (dashed line) at the boundary
of a two-dimensional electronic droplet propagate with opposite
velocities. The quantized Hall conductance for the charge vanishes
while that for the spin is nonvanishing.
        }
\end{center}
\end{figure}

The (global) breaking of time-reversal symmetry is not, strictly speaking,
necessary for integer quantum Hall-like physics. 
As a thought experiment, one can consider, for example,
a noninteracting two-component electronic gas with each component
subjected to a magnetic-like field of equal magnitude
but opposite direction.\cite{Freedman04} 
Each (independent) component is then characterized by its
quantized Hall conductance. The arithmetic average of the two
quantized Hall conductances vanishes 
while their difference is quantized in units of $2e^{*2}/h$ with
$e^{*}$ the effective conserved charge. 
Bernevig and Zhang in  Ref.\ \onlinecite{Bernevig05}
suggested along these lines that,
for some semiconductors with time-reversal symmetric noninteracting 
Hamiltonians, the role of the magnetic field is played
by the intrinsic spin-orbit coupling while the quantum number 
that distinguishes the two components of 
the two-dimensional noninteracting electronic gas is 
the electronic spin.%
\cite{Bernevig05}
If so, the quantized Hall conductance for the electric charge 
(arithmetic average) vanishes while the quantized Hall conductance 
for the spin (difference) is nonvanishing (see Fig.\ \ref{fig: qshe}).

In the proposal of Bernevig and Zhang,
independent quantization of the Hall conductance for each spin 
requires two independent $U(1)$ conserved currents. 
The first one follows from charge conservation. The second one
follows from conservation of the spin quantum number 
perpendicular to the interface in which the electrons are confined.
However, while the intrinsic
spin-orbit coupling breaks the $SU(2)$ spin symmetry down to
its $U(1)$ subgroup, 
the underlying symmetry responsible for the quantization 
of the spin-Hall conductance in Ref.\ \onlinecite{Bernevig05}, 
other spin-orbit couplings such as the Rashba spin-orbit coupling
break this leftover $U(1)$ spin symmetry. This is not to say that  
an unquantized quantum spin Hall effect cannot be present 
if the counterpropagating edge states survive 
the breaking of the residual $U(1)$ spin symmetry. However,
a physical mechanism different from the one protecting the
integer quantum Hall effect must then be invoked for these edge states 
to be robust against (weak) disorder.%
\cite{Zirnbauer92,Brouwer96,Suzuura02a,Suzuura02b,Takane04,Kane05a,%
      Kane05b,Sheng05,Cenke06,CongjunWu06,Sheng06}

\begin{figure}
  \begin{center}
  \includegraphics[width=0.19\textwidth]{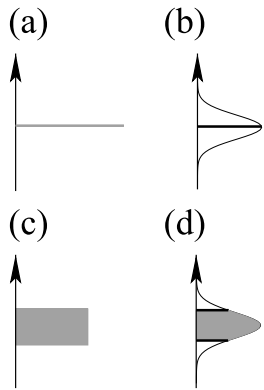}
\caption{
\label{fig: Kane and Mele DOS}
Qualitative plot of the bulk single-particle density of state (DOS)
as a function of the chemical potential for
(a) the integer quantum Hall effect without disorder,
(b) the integer quantum Hall effect with weak disorder,
(c) the $\mathbb{Z}^{\ }_{2}$ quantum spin-Hall state without disorder,
and (d) the $\mathbb{Z}^{\ }_{2}$ quantum spin-Hall state with weak disorder.
The chemical potential runs along the vertical axis 
while the DOS runs along the horizontal axis. 
The gray and white in these figures denote extended and
localized states, respectively.
The black straight lines in (b) and (d) denote the critical energies
at which a quantum phase transition takes place
between two Hall insulating states for (b) 
and between a metallic and an insulating state for (d).
        }
\end{center}
\end{figure}

Kane and Mele showed in 
Refs.\ \onlinecite{Kane05a} and \onlinecite{Kane05b} 
that a noninteracting tight-binding Hamiltonian inspired from graphene, 
with a staggered chemical potential and
with translation invariant intrinsic and 
extrinsic (Rashba) spin-orbit couplings,
realizes a time-reversal symmetric insulating state that they dubbed the
$\mathbb{Z}^{\ }_{2}$ quantum spin-Hall state, provided the chemical potential
lies within the bulk spectral gap (see Fig.\ \ref{fig: Kane and Mele DOS}).
Although the $SU(2)$ spin symmetry is completely broken 
in most of coupling space, parameter space can nevertheless be divided into 
two regions depending on whether the number of Kramers doublet localized 
at the edges in a strip geometry is even or odd. The dispersion
of one Kramers doublet edge state must necessarily
cross the gap in the bulk of the sample
when the number of Kramers doublet edge state is odd, 
in which case it supports an intrinsic quantum spin-Hall effect:
an electric field induces a spin accumulation on the edges transverse 
to the direction of the electric field. 
This insulating state with an odd number of Kramers edge state defines
the $\mathbb{Z}^{\ }_{2}$ quantum spin-Hall state.
It displays a topological texture different from that of the 
integer quantum Hall state.%
\cite{Kane05b,Sheng06,Roy06,Moore06}
The insulating state with
an even number of Kramers edge state is a conventional Bloch insulator.

The effect of disorder is to fill the gap in the bulk spectrum of the 
(clean) $\mathbb{Z}^{\ }_{2}$ quantum spin-Hall state. 
Sufficiently strong disorder is expected to
wash out the $\mathbb{Z}^{\ }_{2}$ quantum spin-Hall state
by removing the edge states
very much in the same way as 
strong disorder does in the integer quantum Hall effect.
On the other hand, Kane and Mele have argued that 
the $\mathbb{Z}^{\ }_{2}$ quantum spin-Hall state
is robust to a weak time-reversal symmetric disorder,
as a single Kramers doublet cannot
undergo back-scattering by a time-reversal symmetric impurity. 
Both expectations were confirmed by a numerical study of 
(i) the four-probe Landauer-B\"uttiker and Kubo formula\cite{Sheng05} 
and of (ii) the spectral flow induced by changes in the twisted
boundary conditions.\cite{Sheng06}
By appealing to the symplectic symmetry of the Kane-Mele Hamiltonian, 
tuning the value of the chemical potential away from the
tails of the disorder-broaden bands towards their center 
should trigger a disorder-induced transition to a metallic state
(see the two mobility edges below and above the metallic state in
Fig.\ \ref{fig: Kane and Mele DOS}).\cite{Hikami80}
Onoda, Avishai, and Nagaosa have raised the 
possibility that the topological nature of the insulating phase might affect
critical properties at this transition.\cite{Onoda06}
Using standard techniques\cite{MacKinnon83} 
to investigate the existence of mobility
edges in tight-binding Hamiltonian perturbed by on-site disorder
(here the Kane-Mele Hamiltonian with random on-site energies distributed
with a box distribution), 
Onoda et al.\ deduced the existence of a mobility edge
separating the $\mathbb{Z}^{\ }_{2}$ quantum spin-Hall state from a metallic
state characterized by the exponent $\nu\approx 1.6$ for the diverging
localization length. This exponent is different from the value
$2.5\lesssim\nu\lesssim2.8$ \cite{Merkt98,Asada02} 
that characterizes the conventional
mobility edge in the two-dimensional symplectic universality class.

\begin{figure}
\begin{center}
\includegraphics[width=6cm]{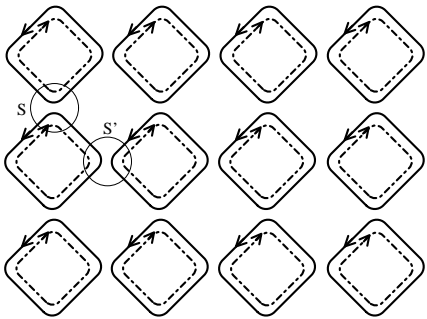}
\end{center}
\caption{
\label{fig: def network model}
A network model is a collection of equipotential lines 
(of the disorder potential) and nodes.
Equipotential lines are closed in the bulk and possibly open at the boundaries.
In this paper, equipotential lines 
are the boundaries of identical squares with rounded corners
while nodes are the midpoints between adjacent rounded corners.
Edge states propagate anticlockwise (full lines)
or clockwise (dashed lines)
along equipotential lines if their spin is up or down, respectively.
Each pair of edge state along an equipotential line
can be arranged into a single Kramers doublet.
Edge states belonging two two different equipotential 
line can exchange their momenta or spin at the nodes of the
network such as $\textsf{S}$ or $\textsf{S}'$.
Each node is thus assigned a $4\times4$ unitary
scattering matrix. 
Two independent copies of the Chalker-Coddington 
network model are obtained in the limit in which
all the $4\times4$ scattering matrices are diagonal
with respect to the spin degrees of freedom.
        } 
\end{figure}

Critical properties at the plateau transition 
in the integer quantum Hall effect
are the same for two very different microscopic models.
There is the effective tight-binding model with random off-diagonal
matrix elements in the basis of Landau wave functions describing
the lowest Landau level.\cite{Huckestein90,Huckestein95}
There is the Chalker-Coddington network model valid for disorder
potentials that vary smoothly on the scale of the cyclotron length.%
\cite{Chalker88}
This agreement supports the notion that, 
for the problem of Anderson localization,
disorder-induced continuous quantum phase transitions fall into 
universality classes determined by dimensionality,
intrinsic symmetry, and topology. 
Furthermore, some network models have provided useful theoretical insights
into the problem of Anderson localization and some have even been 
tractable analytically.\cite{Kramer05} 
The purpose of this paper is to construct a network model 
that realizes a quantum critical point separating
the $\mathbb{Z}^{\ }_{2}$ quantum spin-Hall state from a metallic state.
From this point of view, the network model 
for the two-dimensional symplectic universality class 
introduced in Ref.\ \onlinecite{Merkt98} 
is unsatisfactory as it is built from an even number (two)
of Kramers doublets propagating along the links of the network.
Instead, the network model that we define in Sec.\ \ref{sec: Definition}
has a single Kramers doublet propagating along the links of the
network. Spin is a good quantum number along the links of the network 
so that the spin-up and spin-down components of the Kramers doublet
can be assigned opposite velocities (chiralities).
Scattering takes place at the nodes of the network.
If the scattering matrix is diagonal in spin space, 
the network model realizes the proposal of Bernevig and Zhang:
two copies of the Chalker-Coddington network model
for the integer quantum Hall effect arranged so as not to
break (global) time-reversal symmetry (see Fig.\ \ref{fig: qshe}).
However, we will only demand that the scattering matrix at a node
respects time-reversal symmetry, i.e., 
it can completely break spin-rotation symmetry.
Randomness is introduced through a spin-independent
$U(1)$ random phase along the links.
We also treat the cases of random and non-random 
scattering matrices at the nodes.
In either cases, our spin-filtered chiral network model
captures a continuous quantum phase transition between
the $\mathbb{Z}^{\ }_{2}$ quantum spin-Hall state and the  
metallic state. We find in Sec.\ \ref{sec: Numerics}
the scaling exponent $\nu\approx 2.7$ for
the localization length that is different from the 
exponent $\nu\approx 1.6$ seen by Onoda et al.\ but agrees
with the conventional scaling exponent in the two-dimensional
symplectic universality class.

\section{Definition}
\label{sec: Definition}

To represent the effect of static disorder on the coherent propagation
of electronic waves constrained to a two-dimensional plane and subject
to a strong magnetic field perpendicular to it, 
Chalker and Coddington introduced a 
chiral network model in Ref.\ \onlinecite{Chalker88}.
The Chalker-Coddington network model makes three assumptions.
The disorder is smooth relative to the characteristic microscopic scale: 
the cyclotron length. Equipotential lines of the disorder potential
define the boundaries of mesoscopic quantum Hall droplets along which chiral
edge states propagate coherently. Edge states belonging to distinct 
equipotential lines can only undergo a unitary scattering process by which 
momenta is exchanged provided the distance between the two 
equipotential lines is of order of the cyclotron length.
Such instances are called nodes of the network model.

We are seeking a network model that describes coherent propagation of
electronic waves in a random medium that preserves time-reversal symmetry 
but breaks spin-rotation symmetry, in short a symplectic network model.
A second condition is that the number of edge states that propagate along
equipotential lines can be arranged into an odd number of Kramers doublet.
We choose the number of Kramers doublet to be one for simplicity.
A third condition is that the symplectic network model 
reduces to two independent Chalker-Coddington models 
in some region of parameter space.
The symplectic network model from Ref.\ \onlinecite{Merkt98}
does not fulfill the last two conditions.

\begin{figure}[t]
  \begin{center}
  \includegraphics[width=6cm,clip]{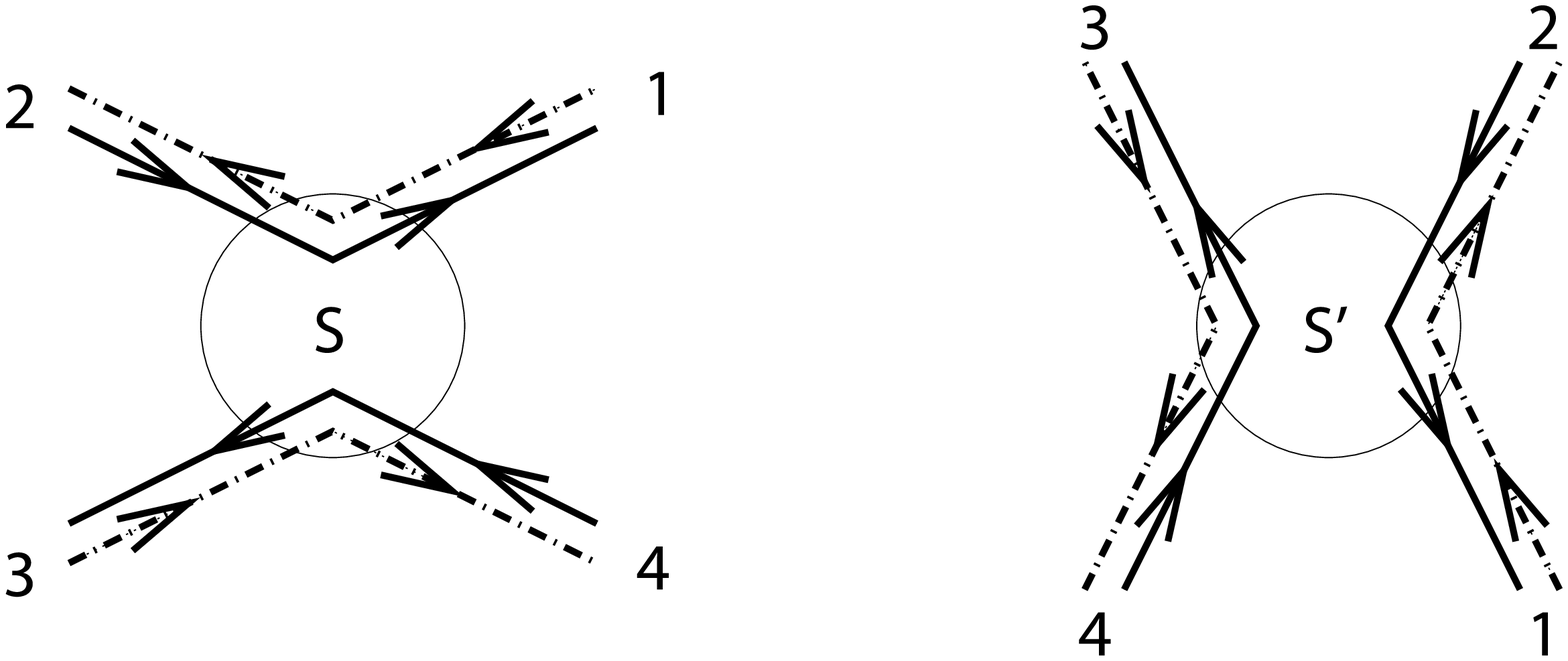} 
\caption{
\label{fig: inequivalent nodes}
There are two nonequivalent nodes
$\mathsf{S}$ and $\mathsf{S}'$
from the point of view of the transfer matrix.
Spin-up (full line)
and spin down (dashed line)
are good quantum numbers on the links
but need not be conserved by the scattering at the nodes.
        }
\end{center}
\end{figure}

Given the last condition, it is natural to start with 
spin-filtered edge states moving along equipotential lines
of the disorder potential depicted as squares with rounded
corners as is done in Fig.\ \ref{fig: def network model}.
The third condition on the symplectic network model
is then satisfied when all the $4\times4$ unitary 
scattering matrices at nodes of the network 
do not couple edge states represented by the arrows along
the full lines with edge states represented by the 
arrows along the dashed lines in 
Fig.\ \ref{fig: def network model}. 
The first two conditions are
otherwise satisfied when all the scattering matrices at the nodes
of the network model from Fig.\ \ref{fig: def network model}
are the most general $4\times4$ unitary matrices 
that respect time-reversal symmetry.
Without loss of generality we choose
a node of type $\mathsf{S}$
from Fig.\ \ref{fig: def network model}.
The most general $4\times4$ unitary scattering matrix
that respects time-reversal symmetry is given by
\begin{equation}
\begin{split}
&
\begin{pmatrix}
\psi^{(\mathrm{o})}_{1\uparrow}
\\
\psi^{(\mathrm{o})}_{2\downarrow}
\\
\psi^{(\mathrm{o})}_{3\uparrow}
\\
\psi^{(\mathrm{o})}_{4\downarrow}
\end{pmatrix}
=
S
\begin{pmatrix}
\psi^{(\mathrm{i})}_{2\uparrow}
\\
\psi^{(\mathrm{i})}_{1\downarrow}
\\
\psi^{(\mathrm{i})}_{4\uparrow}
\\
\psi^{(\mathrm{i})}_{3\downarrow}
\end{pmatrix},
\\
&
S=
\left(
\begin{array}{cc}
 r \sigma^{\ }_0 
& 
t Q 
\\
-t Q^{\dag} 
& 
r \sigma^{\ }_0
\end{array}
\right), \ 
\left(
\begin{array}{cc}
\sigma^{\ }_{2}
& 
0
\\
0
& 
\sigma^{\ }_{2}
\end{array}
\right)
S^{*}
\left(
\begin{array}{cc}
\sigma^{\ }_{2}
& 
0
\\
0
& 
\sigma^{\ }_{2}
\end{array}
\right)=S^\dagger,
\\&
r=\tanh x,
\qquad
t=\frac{1}{\cosh x},
\\
&
Q=
{i}
 \sigma^{\ }_0 \cos\theta \sin\varphi^{\ }_{1}
+
 \sigma^{\ }_1 \sin\theta \cos\varphi^{\ }_{2}
\\
&\hphantom{Q=}
-
 \sigma^{\ }_2 \sin\theta \sin\varphi^{\ }_{2}
+
 \sigma^{\ }_3 \cos\theta \cos\varphi^{\ }_{1},
\end{split}
\label{eq: para S matrix}
\end{equation}
with the labeling of incoming and outgoing
scattering states given in Fig.\ \ref{fig: inequivalent nodes}.
The structure displayed by
Eq.~(\ref{eq: para S matrix}) 
can be understood as follows.
First, the amplitude for an incoming 
spin-filtered edge state not to tunnel must be spin-independent
and thus parametrized by the single real number $r$. Second,
the strength of quantum tunneling at a node can be parametrized
by the positive-valued transmission amplitude $t$ that multiplies 
the purely imaginary quaternion $Q$. 
The purely imaginary quaternion acts on the spin-1/2 degrees of freedom
through the unit $2\times2$ matrix $\sigma^{\ }_{0}$ and the
Pauli matrices $\sigma^{\ }_{1,2,3}$ and must therefore depend
on four real parameters. Third, the local gauge transformation 
$S\to U^{\dag}\,S\,U$ with
$
U=\mathrm{diag}
\left(
\begin{array}{cccc}
e^{+{i}\phi^{\ }_{1}} &
e^{-{i}\phi^{\ }_{1}} &
e^{+{i}\phi^{\ }_{2}} &
e^{-{i}\phi^{\ }_{2}}
\end{array}
\right)
$
can absorb the dependence of $Q$ 
on the two independent phase shifts
$0\leq\varphi^{\ }_{1,2}<2\pi$
compatible with time-reversal symmetry.
At last unitarity delivers the
constraints
$r^{2}+t^{2}=1$
and
$QQ^{\dag}=\sigma^{\ }_{0}$.
Up to an overall sign of $S$ and a local gauge transformation,
$S$ can thus be parametrized by 
\begin{equation}
\big\{(x,\theta)\,\big|\,0\leq x\leq\infty,\qquad 0\leq\theta\leq\pi/2\big\}.
\label{eq: parameter space of network model}
\end{equation}
The boundary $x=\infty$
for which the transmission amplitude vanishes
and the scattering matrix is diagonal
defines the classical limit of the network model.
Quantum tunneling between neighboring plaquettes
in Fig.~\ref{fig: def network model} is very weak
when $x\gg1$. 
In this limit, the network model can be interpreted as follows.
The host $\mathbb{Z}^{\ }_{2}$ quantum spin-Hall state,
i.e., the translation invariant bulk state 
that supports an odd number of Kramers doublet
edge states in a confined geometry free of disorder,
breaks down into droplets of $\mathbb{Z}^{\ }_{2}$ quantum spin-Hall states
separated by smooth random potential barriers.
To appreciate the role played by the parameter
$0\leq\theta\leq\pi/2$, we now consider different values
of $x$ and $\theta$ on the boundary of parameter space.
To this end, it is more convenient to replace the scattering matrix
by two transfer matrices.

Nodes of type $\textsf{S}$ in 
Figs.\ \ref{fig: def network model}
and \ref{fig: inequivalent nodes}
are assigned the transfer matrix
$\widetilde{\mathcal{M}}$,
\begin{widetext}
\begin{subequations}
\begin{equation}
\begin{split}
&
\left(
\begin{array}{c}
\psi^{\ }_{1\uparrow} \\
\psi^{\ }_{1\downarrow} \\
\psi^{\ }_{4\uparrow} \\
\psi^{\ }_{4\downarrow}
\end{array}
\right)
=
\widetilde{\mathcal{M}}
\left(
\begin{array}{c}
\psi^{\ }_{2\uparrow} \\
\psi^{\ }_{2\downarrow} \\
\psi^{\ }_{3\uparrow} \\
\psi^{\ }_{3\downarrow}
\end{array}
\right),
\qquad
\widetilde{\mathcal{M}}=U\,\mathcal{M}\,U^{\dag},
\\
&
U=
\mathrm{diag}
\left(
\begin{array}{cccc}
e^{+\frac{i}{2}(\varphi^{\ }_1+\varphi^{\ }_2)} &
e^{-\frac{i}{2}(\varphi^{\ }_1+\varphi^{\ }_2)} &
e^{-\frac{i}{2}(\varphi^{\ }_1-\varphi^{\ }_2)} &
e^{+\frac{i}{2}(\varphi^{\ }_1-\varphi^{\ }_2)}
\end{array}
\right),
\\
&
\mathcal{M}=
\frac{2}{\cosh2x-\cos2\theta}
\begin{pmatrix}
\sinh x\ \cosh x 
& 
\sin\theta\ \cos\theta
& 
\sinh x\ \cos\theta 
& 
\cosh x\ \sin\theta
\\
-\sin\theta\ \cos\theta 
& 
\sinh x\ \cosh x
& 
-\cosh x\ \sin\theta
& 
\sinh x\ \cos\theta 
\\
\sinh x\ \cos\theta 
& 
\cosh x\ \sin\theta 
& 
\sinh x\ \cosh x 
& 
\sin\theta\ \cos\theta 
\\
-\cosh x\ \sin\theta 
& 
\sinh x\ \cos\theta
& 
-\sin\theta\ \cos\theta 
& 
\sinh x\ \cosh x
\end{pmatrix}.
\end{split}
\label{eq: def mathcalM}
\end{equation}
Nodes of type $\textsf{S}'$ in Fig.\ \ref{fig: def network model}
are assigned the transfer matrix
$\widetilde{\mathcal{M}}'$,
\begin{equation}
\begin{split}
&
\left(
\begin{array}{c}
\psi^{\ }_{2\uparrow}   \\
\psi^{\ }_{2\downarrow} \\
\psi^{\ }_{1\uparrow}   \\
\psi^{\ }_{1\downarrow}
\end{array}
\right)
=
\widetilde{\mathcal{M}}'
\left(
\begin{array}{c}
\psi^{\ }_{3\uparrow}   \\
\psi^{\ }_{3\downarrow} \\
\psi^{\ }_{4\uparrow}   \\
\psi^{\ }_{4\downarrow}
\end{array}
\right),
\qquad
\widetilde{\mathcal{M}}'=
U^{\ }_{1}\,\mathcal{M}'\,U^{\ }_{2},
\\
&
U^{\ }_{1}=
\mathrm{diag}\left(
\begin{array}{cccc}
e^{+\frac{i}{2}(\varphi^{\ }_1+\varphi^{\ }_2)} &
e^{-\frac{i}{2}(\varphi^{\ }_1+\varphi^{\ }_2)} &
e^{+\frac{i}{2}(\varphi^{\ }_1+\varphi^{\ }_2)} &
e^{-\frac{i}{2}(\varphi^{\ }_1+\varphi^{\ }_2)}
\end{array}
\right),
\\
&
U^{\ }_{2}=
\mathrm{diag}\,
\left(
\begin{array}{cccc}
e^{+\frac{i}{2}(\varphi^{\ }_1-\varphi^{\ }_2)} &
e^{-\frac{i}{2}(\varphi^{\ }_1-\varphi^{\ }_2)} &
e^{+\frac{i}{2}(\varphi^{\ }_1-\varphi^{\ }_2)} &
e^{-\frac{i}{2}(\varphi^{\ }_1-\varphi^{\ }_2)} 
\end{array}
\right),
\\
&
\mathcal{M}'=
\begin{pmatrix}
-\cosh x\ \cos\theta
& 
\sinh x\ \sin\theta
& 
\sinh x\ \cos\theta 
& 
-\cosh x\ \sin\theta
\\
-\sinh x\ \sin\theta
& 
-\cosh x\ \cos\theta
& 
\cosh x\ \sin\theta
& 
\sinh x\ \cos\theta 
\\
-\sinh x\ \cos\theta 
& 
\cosh x\ \sin\theta
& 
\cosh x\ \cos\theta
& 
-\sinh x\ \sin\theta
\\
-\cosh x\ \sin\theta
& 
-\sinh x\ \cos\theta
& 
\sinh x\ \sin\theta
& 
\cosh x\ \cos\theta
\end{pmatrix}.
\end{split}
\label{eq: def mathcalM'}
\end{equation}
\end{subequations}
\end{widetext}
Here, the convention for initial and final 
scattering states is given in Fig.\ \ref{fig: inequivalent nodes}.
One verifies that, for all values of
$0\leq\varphi^{\ }_{1,2}<2\pi$,
$0\leq x\leq\infty$, and $0\leq\theta\leq\pi/2$
that parametrize
$\widetilde{\mathcal{M}}$ and $\widetilde{\mathcal{M}}^{\prime}$,
the conditions for pseudo-unitary 
\begin{equation}
\mathcal{A}
\begin{pmatrix}
+\sigma^{\ }_3 
& 
0 
\\ 
0 
& 
-\sigma^{\ }_3
\end{pmatrix}
\mathcal{A}^\dagger
=
\begin{pmatrix}
+\sigma^{\ }_3 
& 
0 
\\ 
0 
& 
-\sigma^{\ }_3
\end{pmatrix}
\end{equation}
and time-reversal symmetry
\begin{equation}
\begin{pmatrix}
i\sigma^{\ }_2 
& 
0 
\\ 
0 
& 
i\sigma^{\ }_2
\end{pmatrix}
\mathcal{A}
\begin{pmatrix}
-i\sigma^{\ }_2 
& 
0 
\\ 
0 
& 
-i\sigma^{\ }_2
\end{pmatrix}
=
\mathcal{A}^*
\end{equation}
hold ($\mathcal{A}=\mathcal{M}$ or $\mathcal{M}'$). 

\begin{figure}[t]
\begin{center}
\includegraphics[width=6cm]{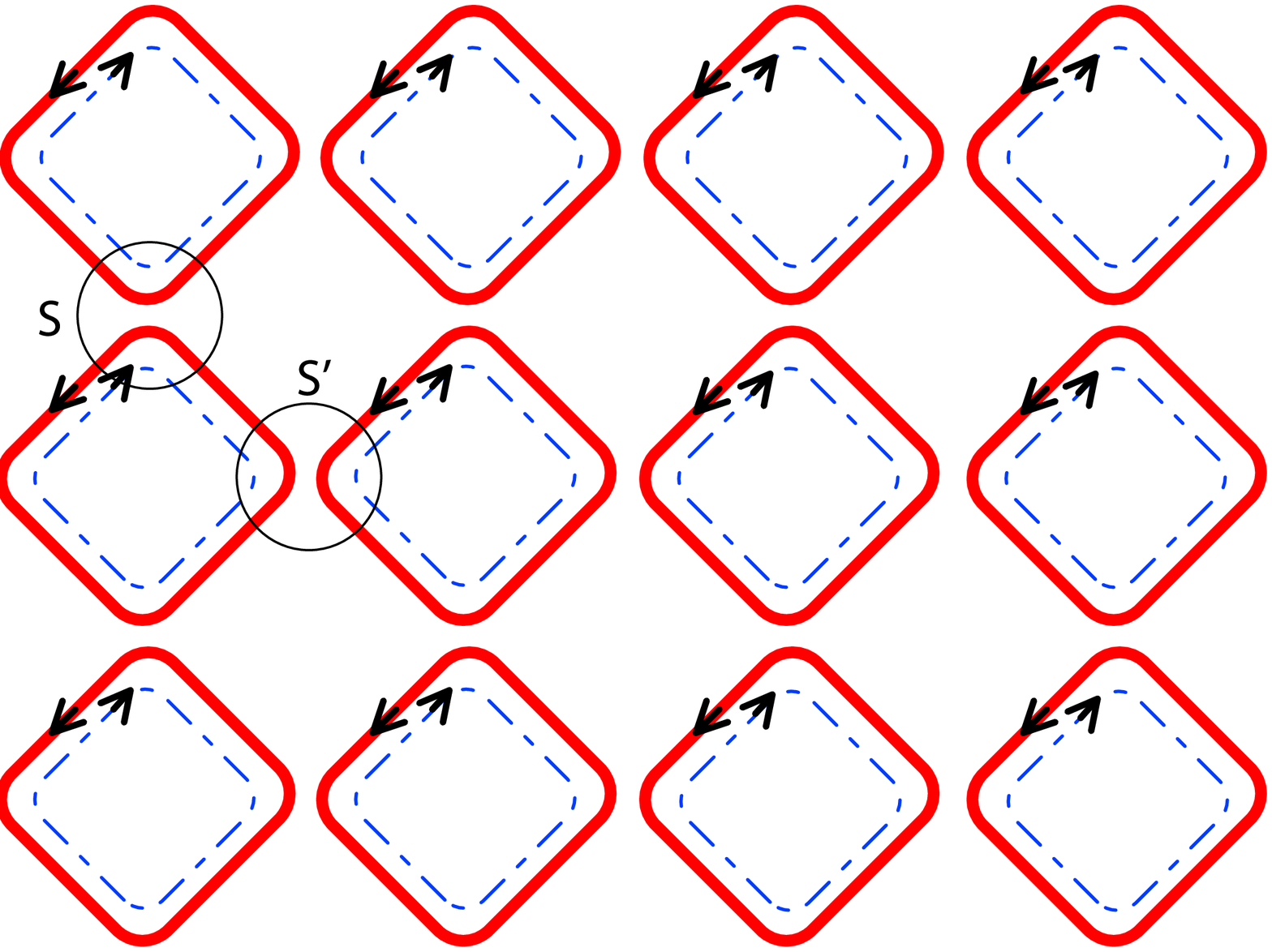}
\end{center}
\caption{
(Color online:)
The network model at $\theta=0$ decouples into 
two networks depicted by the thick (red) and thin (blue) lines.
Full and dashed lines distinguish propagation along the links of the networks 
of up and down spins, respectively.}
\label{fig: network-theta=0}
\end{figure}

\textit{Along the boundary $\theta=0$,}
the transfer matrices (\ref{eq: def mathcalM}) and (\ref{eq: def mathcalM'})
reduce to 
\begin{subequations}
\begin{equation}
\mathcal{M}=
\begin{pmatrix}
\coth x 
& 
0
& 
1/\sinh x 
& 
0
\\
0
& 
\coth x
& 
0
& 
1/\sinh x 
\\
1/\sinh x 
& 
0
& 
\coth x 
& 
0
\\
0
& 
1/\sinh x 
& 
0
& 
\coth x
\end{pmatrix}
\end{equation}
and
\begin{equation}
\mathcal{M}^{\prime}=
\begin{pmatrix}
-\cosh x\ 
& 
0
& 
\sinh x 
& 
0
\\
0
& 
-\cosh x
& 
0
& 
\sinh x 
\\
-\sinh x 
& 
0
& 
\cosh x 
& 
0
\\
0
& 
-\sinh x 
& 
0
& 
\cosh x
\end{pmatrix},
\end{equation}
\end{subequations}
respectively. 
As is depicted in Fig.\ \ref{fig: network-theta=0},
up and down spins have decoupled into two independent
Chalker-Coddington models, 
each of which describes the integer quantum Hall plateau transition.
Whenever $x=0$ or $x=\infty$
either $\mathcal{M}$ or $\mathcal{M}^{\prime}$
is diagonal so that edge states cannot escape the equipotential lines
encircling the local extrema of the disorder potential.
These are strongly insulating phases
characterized by different integer
topological (Chern) numbers;
one for each spin direction.
Across the plateau transition the number of edge states changes 
by one for each spin, and so does the number of Kramers doublet edge mode.
This implies that the two insulating phases are distinct
in the $\mathbb{Z}^{\ }_2$ classification.
Quantum tunneling is strongest at the integer quantum Hall transition
defined by the condition
$\mathcal{M}\sim\mathcal{M}^{\prime}$
for which $x^{\ }_{\mathrm{cc}}=\ln(1+\sqrt{2})$.
(By $\sim$ is meant equality in magnitude of all matrix elements.)

\begin{figure}[t]
\begin{center}
\includegraphics[width=6cm]{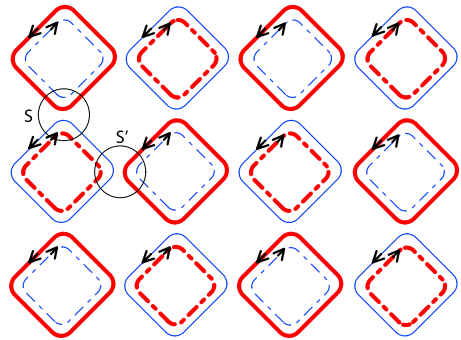}
\end{center}
\caption{
(Color online:)
The network model at $\theta=\pi/2$ decouples into 
two networks depicted by the thick (red) and thin (blue) lines.
Full and dashed lines distinguish propagation along the links of the networks 
of up and down spins, respectively.
        } 
\label{fig: network-theta=pi/2}
\end{figure}

\textit{Along the boundary $\theta=\pi/2$,}
the transfer matrices (\ref{eq: def mathcalM}) and (\ref{eq: def mathcalM'})
reduce to 
\begin{subequations}
\begin{equation}
\mathcal{M}=
\begin{pmatrix}
\tanh x
& 
0
& 
0
& 
1/\cosh x
\\
0
& 
\tanh x
& 
-1/\cosh x 
& 
0
\\
0
& 
1/\cosh x
& 
\tanh x
& 
0
\\
-1/\cosh x
& 
0
& 
0
& 
\tanh x
\end{pmatrix}
\end{equation}
and
\begin{equation}
\mathcal{M}^{\prime}=
\begin{pmatrix}
0
& 
\sinh x
& 
0
& 
-\cosh x 
\\
-\sinh x 
& 
0
& 
\cosh x 
& 
0
\\
0
& 
\cosh x
& 
0
& 
-\sinh x
\\
-\cosh x
& 
0
& 
\sinh x
& 
0
\end{pmatrix},
\end{equation}
\end{subequations}
respectively. 
The network model has decoupled into two independent network models
as is depicted in Fig.\ \ref{fig: network-theta=pi/2}.
The $U(1)$ residual spin-rotation symmetry
at $\theta=0$ is maximally broken by $\theta=\pi/2$.
When $x=\infty$,
$\mathcal{M}$ becomes diagonal while
$\mathcal{M}^{\prime}$ is off-diagonal
so that edge states cannot escape the equipotential lines
encircling the local extrema of the disorder potential.
The point $x=0$ is dominated by quantum tunneling since
$\mathcal{M}\sim\mathcal{M}^{\prime}$
are then both anti-diagonal. 
(By $\sim$ is meant equality in magnitude of all matrix elements.)
Furthermore,
at $x=0$, propagation of Kramers doublets is ballistic along
decoupled one-dimensional chiral channels.
Each network model depicted in Fig.\ \ref{fig: network-theta=pi/2}
belongs to the unitary universality class (without topological term)
when $\theta=\pi/2$ and $0\leq x\leq\infty$.
We thus anticipate an unstable fixed point at $x=0$ 
describing a metallic phase and an insulating
phase for $x>0$.

\begin{figure}
\begin{center}
\includegraphics[width=6cm]{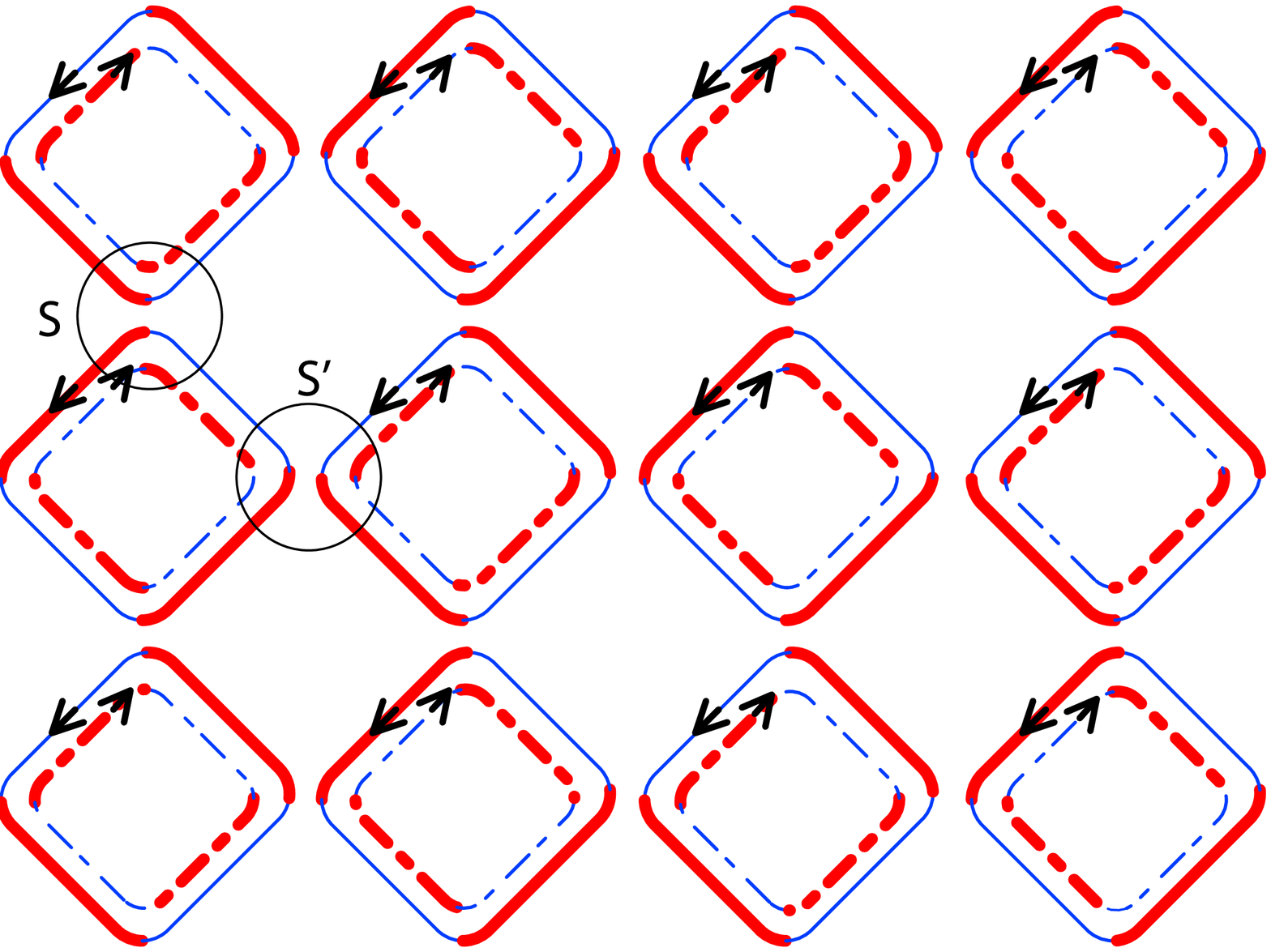}
\end{center}
\caption{
(Color online:)
The network model at $x=0$ decouples into
two networks depicted by the thick (red) and thin (blue) lines.
Full and dashed lines distinguish propagation along the links of the networks 
of up and down spins, respectively.
        } 
\label{fig: network-x=0}
\end{figure}

\textit{Along the boundary $x=0$,}
the transfer matrices (\ref{eq: def mathcalM}) and (\ref{eq: def mathcalM'})
reduce to
\begin{subequations} 
\begin{equation}
\mathcal{M}=
\begin{pmatrix}
0
& 
\cot\theta
& 
0
& 
1/\sin\theta
\\
-\cot\theta 
& 
0
& 
-1/\sin\theta\
& 
0
\\
0
& 
1/\sin\theta 
& 
0
& 
\cot\theta 
\\
-1/\sin\theta 
& 
0
& 
-\cot\theta 
& 
0
\end{pmatrix}
\end{equation}
and
\begin{equation}
\mathcal{M}^{\prime}=
\begin{pmatrix}
-\cos\theta
& 
0
& 
0
& 
-\sin\theta
\\
0
& 
-\cos\theta
& 
\sin\theta
& 
0
\\
0
& 
\sin\theta
& 
\cos\theta
& 
0
\\
-\sin\theta
& 
0
& 
0
& 
\cos\theta
\end{pmatrix},
\end{equation}
\end{subequations}
respectively.
The network model has decoupled into two independent network models
as is depicted in Fig.\ \ref{fig: network-x=0}.
When $\theta=0$,
$\mathcal{M}$ is off-diagonal while
$\mathcal{M}^{\prime}$ is diagonal
yielding a strongly insulating phase.
The point $\theta=\pi/2$ is dominated by quantum tunneling since
$\mathcal{M}\sim\mathcal{M}^{\prime}$
are then both anti-diagonal.
(By $\sim$ is meant equality in magnitude of all matrix elements.)
Furthermore,
at $\theta=\pi/2$, propagation of Kramers doublets is ballistic along
decoupled one-dimensional chiral channels. 
Each network model depicted in Fig.\ \ref{fig: network-x=0}
belongs to the unitary universality class (without topological term)
when $x=0$ and $0\leq\theta\leq\pi/2$.
We thus anticipate an unstable fixed point at $\theta=\pi/2$ 
describing a metallic phase and an insulating
phase for $0\leq\theta<\pi/2$.

Observe that the duality relation
\begin{subequations}
\label{eq: duality between x=0 and theta=piover2}
\begin{equation}
\tanh x= \cos\theta
\end{equation}
implies that
\begin{equation}
\begin{split}
&
\mathcal{M} (x=0,\theta)\sim
\mathcal{M}'(x,\theta=\pi/2),
\\
&
\mathcal{M}'(x=0,\theta)\sim
\mathcal{M} (x,\theta=\pi/2).
\end{split}
\end{equation}
\end{subequations}
(By $\sim$ is meant equality in magnitude of all matrix elements.)

\begin{figure}[b]
\begin{center}
\includegraphics[width=6cm]{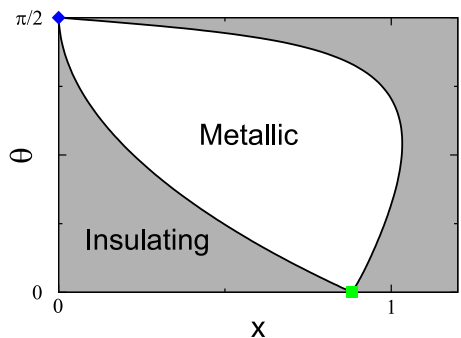}
\end{center}
\caption{
(Color online:)
Expected phase diagram from the analysis
of the network model along the boundaries of parameter space%
~(\ref{eq: parameter space of network model}).
The fixed point denoted by a filled (green) square 
along the boundary $\theta=0$ is the unstable quantum critical 
point located at $x^{\ }_{\mathrm{cc}}=\ln(1+\sqrt{2})$
separating two insulating phases in the
Chalker-Coddington model.
The fixed point denoted by the filled (blue) rhombus 
at the upper left corner is the unstable metallic phase.
The shape of the metallic phase is controlled by the
symmetry crossover between the unitary and symplectic
symmetry classes. 
        } 
\label{fig: predicted phase diagram}
\end{figure}

From the analysis of the network model
on the boundaries of parameter space, 
we deduce the qualitative phase diagram
shown in Fig.\ \ref{fig: predicted phase diagram}.
The numerics of Sec.\ \ref{sec: Numerics}
confirm the overall topology of this phase diagram.

\begin{figure*}
  \begin{center}
  \includegraphics[width=14cm,clip]{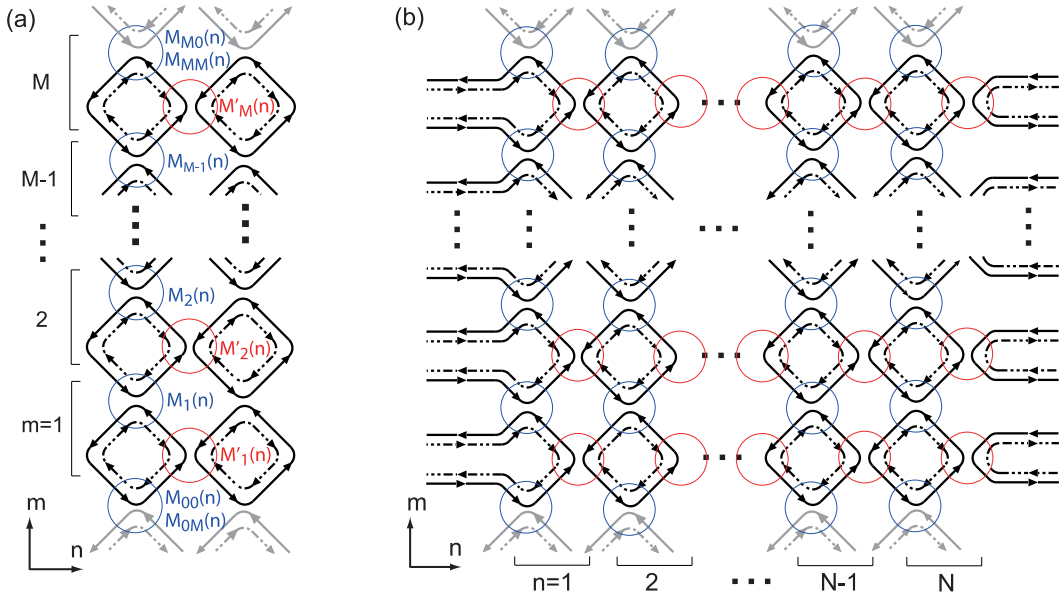}
\caption{
\label{fig: slice and wire}
(Color online:)
(a) A slice of the network model assuming periodic boundary conditions
in the transverse direction, here represented by the gray links. 
(b) Wire geometry of the network model.
        }
\end{center}
\end{figure*}

The definition of the 
two-dimensional spin-filtered chiral network model for the 
$\mathbb{Z}^{\ }_2$ quantum spin-Hall effect
is completed by specifying the boundary conditions.
These are dictated by the numerical method that we shall use
in Sec.\ \ref{sec: Numerics}. Following MacKinnon and Kramer,
we seek the transfer matrix of a long but narrow sample connected 
at both ends to semi-infinite ideal metallic leads.
To minimize finite size effects, 
we impose periodic boundary conditions
in the transverse direction.
The transfer matrix is then a
$4M\times4M$ pseudo-unitary matrix
that maps $4M$ plane waves from the left lead 
into $4M$ plane waves from the right lead
that we define as follows. 
First, we consider a slice of the sample 
that we label by the integer $n=1,2,\ldots,N$
as is depicted in Fig.\ \ref{fig: slice and wire}a
($N\gg 4M$).
We assign to this slice the $4M\times4M$ pseudo-unitary matrix
$\mathcal{M}^{\ }_{\mathrm{sl}}(n)$,
\begin{subequations}
\label{eq: def trsf matrix PBC}
\begin{widetext}
\begin{equation}
\begin{split}
&
\mathcal{M}^{\ }_{\mathrm{sl}}(n):=
\mathcal{M}^{\ }_{\mathsf{S}'}(n)\
U^{(2)}_{\mathrm{sl}}(n)\
\mathcal{M}^{\ }_{\mathsf{S} }(n)\
U^{(1)}_{\mathrm{sl}}(n),
\\
&
U^{(1)}_{\mathrm{sl}}(n)=
\mathrm{diag}
\begin{pmatrix}
e^{+i\phi^{(1)}_{1}(n)}
&
e^{-i\phi^{(1)}_{1}(n)}
&
\cdots
&
e^{+i\phi^{(1)}_{2M}(n)}
&
e^{-i\phi^{(1)}_{2M}(n)}
\end{pmatrix},
\\
&
\mathcal{M}^{\ }_{\mathsf{S} }(n)=
\begin{pmatrix}
\mathcal{M}^{\ }_{00}(n) 
& 
0
& 
\cdots
&
0 
& 
\mathcal{M}^{\ }_{0M}(n) 
\\
0
& 
\mathcal{M}^{\ }_1(n) 
&
0
&
\cdots
&
0
\\
\vdots
& 
\ddots
& 
\ddots 
&
\ddots
&
\vdots
\\
0
& 
\cdots
& 
0
& 
\mathcal{M}^{\ }_{M-1}(n) 
&
0
\\
\mathcal{M}^{\ }_{M0}(n) 
&
0 
& 
\cdots
&
0 
& 
\mathcal{M}^{\ }_{MM}(n)
\end{pmatrix},
\\
&
U^{(2)}_{\mathrm{sl}}(n)=
\mathrm{diag}
\begin{pmatrix}
e^{+i\phi^{(2)}_{1}(n)}
&
e^{-i\phi^{(2)}_{1}(n)}
&
\cdots
&
e^{+i\phi^{(2)}_{2M}(n)}
&
e^{-i\phi^{(2)}_{2M}(n)}
\end{pmatrix},
\\
&
\mathcal{M}^{\ }_{\mathsf{S}'}(n)=
\mathrm{diag}
\begin{pmatrix}
\mathcal{M}'_{1}(n)
&
\cdots
&
\mathcal{M}'_{M}(n)
\end{pmatrix}.
\end{split}
\end{equation}
Here, $\mathcal{M}^{\ }_{m}(n) $ with $m=1,\ldots,M-1$
and $\mathcal{M}'_{m}(n) $ with $m=1,\ldots,M$
are given by Eqs.\ (\ref{eq: def mathcalM}) and (\ref{eq: def mathcalM'}),
respectively, while we have imposed periodic boundary conditions 
in the transverse direction with the choice
\begin{align}
&
\mathcal{M}_{00}(n)=
\mathcal{M}_{MM}(n)=
\frac{2}{\cosh 2x - \cos 2\theta}
\begin{pmatrix}
\sinh x \cosh x 
& 
\sin\theta \cos\theta 
\\
-\sin\theta \cos\theta 
& 
\sinh x \cosh x
\end{pmatrix},
\\
&
\mathcal{M}_{0M}(n)=
\mathcal{M}_{M0}(n)=
\frac{2}{\cosh 2x - \cos 2\theta}
\begin{pmatrix}
\sinh x \cos\theta 
& 
\sin\theta \cosh x 
\\
-\sin\theta \cosh x 
& 
\sinh x \cos\theta
\end{pmatrix}.
\end{align}
\end{widetext}
The phases $\phi^{(l)}_{m}(n)$ with
$l=1,2$, $m=1,\ldots,2M$, and $n=1,\ldots,N$ take values
between $0$ and $2\pi$.
Second, we assign to the 
quasi-one dimensional network model depicted in
Fig.\ \ref{fig: slice and wire}b the transfer matrix
\begin{equation}
\mathcal{M}^{\ }_{\mathrm{tot}}:=
\prod_{n=1}^{N}
\mathcal{M}^{\ }_{\mathrm{sl}}(n).
\end{equation}
\end{subequations}
This completes the definition of the 
two-dimensional spin-filtered chiral network model for the 
$\mathbb{Z}^{\ }_2$ quantum spin-Hall effect.

We close Sec.\ \ref{sec: Definition}
by showing that $\mathcal{M}^{\ }_{\mathrm{tot}}$
belongs to the Lie group $SO^{*}(4M)$. By construction,
flux conservation,
\begin{equation}
\mathcal{M}^{\ }_{\mathrm{tot}}\,
\Sigma_3\,
\mathcal{M}^{\dag}_{\mathrm{tot}}
=\Sigma^{\ }_3,
\qquad
\Sigma^{\ }_3=
\begin{pmatrix}
\sigma^{\ }_3 
& 
0 
\\ 
0 
& 
-\sigma^{\ }_3
\end{pmatrix}
\otimes
I^{\ }_M,
\label{eq: punitary}
\end{equation}
and time-reversal symmetry,
\begin{equation}
\Sigma^{\ }_2\,
\mathcal{M}^{\ }_{\mathrm{tot}}\,
\Sigma^{T}_2
=
\mathcal{M}^{*}_{\mathrm{tot}},
\qquad
\Sigma^{\ }_2=
i\sigma^{\ }_2\otimes I^{\ }_{2M},
\label{eq: TRS}
\end{equation}
where $I^{\ }_{2M}$ is the $2M\times 2M$ unit matrix hold.
It follows from Eq.\ (\ref{eq: TRS}) that
\begin{equation}
\mathcal{M}^{\dag}_{\mathrm{tot}}=
\Sigma^{\ }_2\,
\mathcal{M}^{T}_{\mathrm{tot}}\,
\Sigma^{T}_2.
\label{eq: step 1}
\end{equation}
Substituting Eq.~(\ref{eq: step 1})
into Eq.\ (\ref{eq: punitary}) yields
\begin{equation}
\begin{split}
&
\mathcal{M}^{\ }_{\mathrm{tot}}\,
\Sigma^{\ }_1\,
\mathcal{M}^{T}_{\mathrm{tot}}
=
\Sigma^{\ }_1,
\\
&
\Sigma^{\ }_1=\Sigma^{\ }_3\Sigma^{\ }_2=
\begin{pmatrix}
\sigma^{\ }_1 
& 
0 
\\ 
0 
& 
-\sigma^{\ }_1
\end{pmatrix}
\otimes
I^{\ }_M.
\end{split}
\label{eq: Sigma_1}
\end{equation}
We introduce the matrix
\begin{equation}
A=\frac{1}{\sqrt2}
\begin{pmatrix}
\sigma^{\ }_2+\sigma^{\ }_3 
& 
0 
\\ 
0 
& 
\sigma^{\ }_2-\sigma^{\ }_3
\end{pmatrix}
\otimes
I^{\ }_M,
\end{equation}
and write $\Sigma^{\ }_1=-iA A^T$.
Equation (\ref{eq: Sigma_1}) then reads
\begin{equation}
\widehat{\mathcal{M}}^{\ }_{\mathrm{tot}}\,
\widehat{\mathcal{M}}^{T }_{\mathrm{tot}}
=1,
\qquad
\widehat{\mathcal{M}}^{\ }_{\mathrm{tot}}
=A\,\mathcal{M}^{\ }_{\mathrm{tot}}\,A,
\label{eq: orthogonality}
\end{equation}
where we have used the identity $A^2=1$.
We can rewrite Eq.\ (\ref{eq: punitary}) in terms of
$\widehat{\mathcal{M}}^{\ }_{\mathrm{tot}}$,
\begin{equation}
\begin{split}
&
\widehat{\mathcal{M}}^{\ }_{\mathrm{tot}}\,
\Sigma^{\ }_2\,
\widehat{\mathcal{M}}^{\dag}_{\mathrm{tot}}
=
\Sigma^{\ }_2,
\\
&
iA\,\Sigma^{\ }_3\,A
=
\begin{pmatrix}
i\sigma^{\ }_2 
& 
0 
\\ 
0 
& 
i\sigma^{\ }_2
\end{pmatrix}
\otimes
I^{\ }_M
=
\Sigma^{\ }_2.
\end{split}
\label{eq: anti-unitary}
\end{equation}
With an orthogonal transformation that exchanges rows and
columns, we can bring $\Sigma^{\ }_2$ into the form,
\begin{equation}
O\,\Sigma^{\ }_2\,O^T=
\begin{pmatrix}
0 
& 
I^{\ }_{2M} 
\\ 
-I^{\ }_{2M} 
& 
0
\end{pmatrix}
=J^{\ }_{2M}.
\end{equation}
We thus conclude that 
$O\,\widehat{\mathcal{M}}^{\ }_{\mathrm{tot}}\,O^T$
is an element of the group $SO^*(4M)$ 
defined by the conditions,
\begin{equation}
g\,J^{\ }_{2M}\,g^{\dag}=J^{\ }_{2M},
\quad
g\, g^{T}=I^{\ }_{4M},
\quad
g\in GL(4M,\mathbb{C}).
\end{equation}

\section{Numerics}
\label{sec: Numerics}

This section is devoted to a numerical study of
the dependence of the smallest Lyapunov exponent
of the transfer matrix
$\mathcal{M}^{\   }_{\mathrm{tot}}$ 
defined in Eq.~(\ref{eq: def trsf matrix PBC}),
as a function of the width $M$ of the 
quasi-one dimensional network model.
Although $\mathcal{M}^{\   }_{\mathrm{tot}}$
is taken from a statistical ensemble that we will
specify below, Lyapunov exponents are 
self-averaging random variables
for an infinitely long quasi-one dimensional network model, 
$N\to\infty$.%
\cite{Johnston83,Brouwer96}

The eigenvalues of the $4M\times4M$ Hermitian matrix
$\mathcal{M}^{\dag}_\mathrm{tot}\,\mathcal{M}^{\   }_\mathrm{tot}$
are doubly degenerate and written as $\exp(\pm 2X_j)$ with
$0<X_1<X_2<\ldots<X_{M}$.
The localization length $\xi^{\ }_M$ is then given by
\begin{equation}
\xi^{\ }_M=\lim_{N\to\infty}\frac{N}{X_1}.
\end{equation}
The localization length $\xi^{\ }_{M}$
is a finite and self-averaging length scale
that controls the exponential decay of the Landauer conductance
for any fixed width $M$ of the infinitely long 
quasi-one dimensional network model, 
as the transfer matrix~(\ref{eq: def trsf matrix PBC})
belongs to the group $SO^{*}(4M)$.%
\cite{Brouwer96}
It is of course impossible to study infinitely long 
quasi-one dimensional network models
numerically and we shall approximate
$\xi^{\ }_{M}$
with $\xi^{\ }_{M,N}$ obtained from the Lyapunov exponents
of a finite but long quasi-one dimensional network model
made of $N$ slices. In our numerics we have set 
$N=5\times10^5\sim8\times10^6$.

As shown by MacKinnon and Kramer,\cite{MacKinnon83}
criticality in two dimensions can be accessed from the
dependence of the normalized localization length
\begin{equation}
\Lambda:=\xi^{\ }_M/M
\end{equation}
on the width $M$ of the quasi-one dimensional network model.
For example, if $\xi$ denotes the two-dimensional localization length
and if $\xi$ diverges according to the power law
\begin{equation}
\xi\sim
|z-z^{\ }_{c}|^{-\nu}
\end{equation}
upon tuning of a single microscopic parameter $z$ close to 
its critical value $z^{\ }_{c}$, 
the singular part of $\Lambda$ as $M\to\infty$
should be given by a scaling function\cite{Slevin99}
\begin{equation}
\Lambda\sim
F(\chi M^{1/\nu},\zeta M^y,\ldots).
\label{eq:scaling-func-irr}
\end{equation}
Here, $\chi$ and $\zeta$ are 
the single relevant and dominant irrelevant scaling variables, respectively.%
\cite{Cardy96}
The largest irrelevant scaling exponent satisfies $y<0$. 
We assume that $F$ can be expanded in powers 
of $\zeta M^y$ and $\chi M^{1/\nu}$,
\begin{equation}
\begin{split}
\Lambda\sim&\,
\sum_{p=0}^{\infty}
\sum_{q=0}^{\infty}
F^{\ }_{p,q}
\zeta^{p}
 \chi^{q}  
M^{py+q/\nu}
\end{split}
\label{eq:scaling-func-irr-truncate}
\end{equation}
where $F^{\ }_{p,q}\in\mathbb{R}$ are the expansion coefficients.
We also assume that the relevant scaling variable $\chi$ 
is linearly related to $|z-z^{\ }_{c}|$
while the irrelevant scaling variable $\zeta$ is a constant
in the vicinity of the critical point.
Finally, for any given $0\leq\theta\leq\pi/2$ from 
the scattering matrix~(\ref{eq: para S matrix}),
we identify the microscopic parameter $z$
as the parameter $0\leq x\leq\infty$.
This motivates the scaling ansatz
\begin{subequations}
\begin{equation}
\begin{split}
\Lambda=&\,
\sum_{q=0}^{3}
f^{(\theta)}_{0,q}
\left(x-x^{(\theta)}_{c}\right)^{q}
M^{q/\nu}
\\
&\,
+
\sum_{q=0}^{2}
f^{(\theta)}_{1,q}
\left(x-x^{(\theta)}_{c}\right)^{q}
M^{y+q/\nu}
\end{split}
\label{eq: scaling ansatz lambda}
\end{equation}
with the 10 real-valued fitting parameters
\begin{equation}
\nu,\
y,\
x^{(\theta)}_{c},\
\Lambda^{(\theta)}_{c}:=
f^{(\theta)}_{0,0},
\end{equation}
and
\begin{equation}
f^{(\theta)}_{0,1},\
f^{(\theta)}_{0,2},\
f^{(\theta)}_{0,3},\
f^{(\theta)}_{1,0},\
f^{(\theta)}_{1,1},\
f^{(\theta)}_{1,2}.
\end{equation}
\end{subequations}
Observe that single-parameter scaling is obeyed by 
\begin{equation}
\Lambda':=
\Lambda
-
\sum_{p=1}^{\infty}
\sum_{q=0}^{\infty}
F^{\ }_{p,q}
\zeta^{p}
 \chi^{q}  
M^{py+q/\nu},
\label{eq:corrected_Lambda}
\end{equation}
or, in practice,
\begin{align}
\Lambda':=&\,
\Lambda
-
\sum_{q=0}^{2}
f^{(\theta)}_{1,q}
\left(x-x^{(\theta)}_{c}\right)^{q}
M^{y+q/\nu}
\nonumber \\
=&\,
\sum^3_{q=0}f^{(\theta)}_{0,q}
\left(x-x^{(\theta)}_{c}\right)^{q}
M^{q/\nu}.
\label{eq: scaling ansatz lambda'}
\end{align}
The values taken by the width $M$ 
of the quasi-one dimensional network model are
$M=4,\, 8,\, 16,\, 32,\, 64$. 
To reduce the statistical error,
average over $16$ different realizations of the disorder potential
are calculated for any given ${M,x,\theta}$
when $\theta$ is not random and ${M,x}$ otherwise.

The disorder potential is modeled in two different ways, i.e., we
introduce disorder in the transfer matrix~(\ref{eq: def trsf matrix PBC})
as follows. Case I: $0\leq\theta\leq\pi/2$ and $0\leq x\leq\infty$
are the same for all nodes and randomness is introduced by 
taking all the phases $\phi^{(l)}_{m}(n)$ with
$l=1,2$, $m=1,\ldots,2M$, and $n=1,\ldots,N$ to be 
independently and uniformly distributed between 
$0$ and $2\pi$.
Case II: 
in addition to the randomness in the phases
$\phi^{(l)}_{m}(n)$ we allow $\theta$ to be independently
distributed with the probability $\sin(2\theta)$
between $0$ and $\pi/2$ at each node of the network.
As before, $0\leq x\leq\infty$ is the same for all nodes.

\begin{figure}[t]
\begin{center}
\includegraphics[width=8cm]{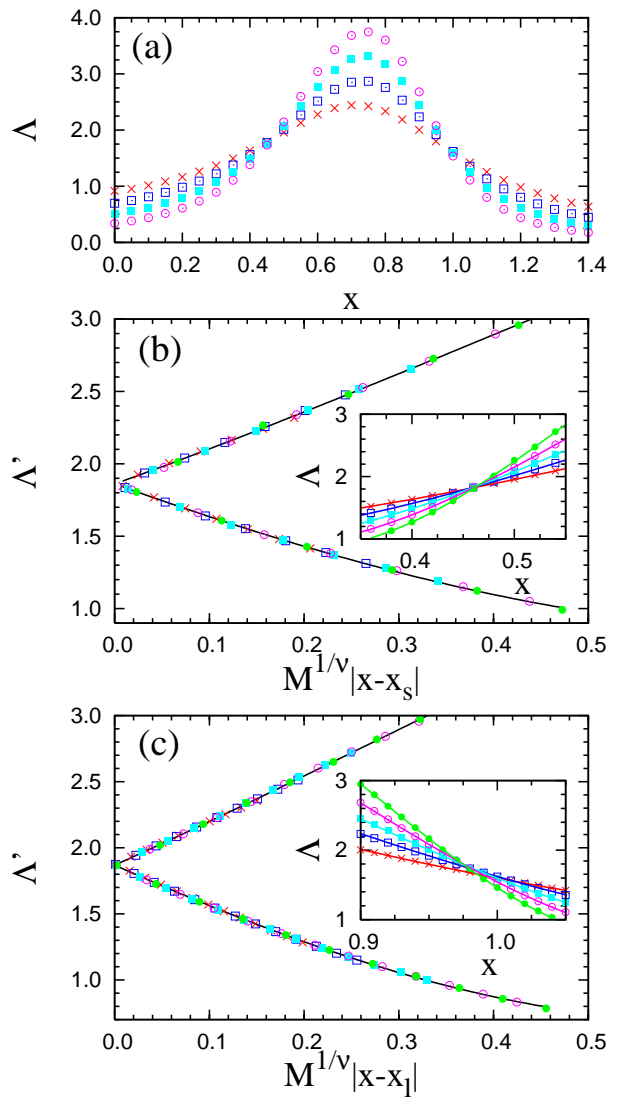}
\end{center}
\caption{
(Color online:)
(a) Normalized localization length $\Lambda$
as a function of $x$ at fixed $\theta=3\pi/16$
for the widths 
$M=4$ (crosses), 
$M=8$ (open squares), 
$M=16$ (filled squares), 
$M=32$ (open circles), 
and $64$ (filled circles)
of the quasi-one dimensional network.
Error bars are much smaller than symbol sizes.
There are two values of $x$ 
($x^{\ }_{s}\approx 0.46$ 
and 
$x^{\ }_{l}\approx  0.97$)
for which $\Lambda$ does not appear to depend on $M$.
(b) A fit of the data shown in (a)
with the help of the one-parameter 
scaling ansatz~(\ref{eq: scaling ansatz lambda'})
when $x$ is close to $x^{\ }_{s}$.
Inset: A blow up of (a) in the vicinity of $x^{\ }_{s}$. 
(c) A fit of the data shown in (a)
with the help of the one-parameter 
scaling ansatz~(\ref{eq: scaling ansatz lambda'})
when $x$ is close to $x^{\ }_{l}$.
Inset: A blow up of (a) in the vicinity of $x^{\ }_{l}$. 
        } 
\label{fig:Lambda_theta=3pi/16}
\end{figure}

\begin{table}[b]
\begin{center}
\caption{
Critical exponent, normalized localization length,
and (minimal) node parameter $x_s$
as a function of $\theta$.
The poor agreement at $\theta=\pi/4$ is probably due to
the presence of a large crossover length scale
near the boundary $x=0$.
        }
\label{tab:xs}
\begin{tabular}{c c c c}
\hline\hline
$\theta$ \quad & $\nu$ \quad & $\Lambda_c$ \quad & $x_s$ \\
\hline
$\pi/8$ \quad & $2.85 \pm 0.30$ \quad  & $1.87\pm0.09$ \quad & $0.667 \pm 0.004$ \\
$3\pi/16$ \quad  & $2.77 \pm 0.16$ \quad & $1.86\pm0.02$ \quad & $0.465 \pm 0.001$ \\
$7\pi/32$ \quad & $2.73 \pm 0.05$ \quad & $1.90\pm0.01$ \quad & $0.244 \pm 0.001$ \\
$\pi/4$ * \quad & $2.17 \pm 0.21$ \quad & $1.82\pm0.01$ \quad & $0.016 \pm 0.001$ \\
\hline\hline
\end{tabular}
\end{center}
\end{table}

\begin{table}[b]
\begin{center}
\caption{
Critical exponent, normalized localization length,
and (maximal) node parameter $x_l$
as a function of $\theta$.
        }
\label{tab:xl}
\begin{tabular}{c c c c}
\hline\hline
$\theta$ \quad & $\nu$ \quad & $\Lambda_c$ \quad & $x_l$ \\
\hline
$\pi/8$ \quad & $2.78 \pm 0.19$ \quad  & $1.94\pm0.10$ \quad & $0.972 \pm 0.004$ \\
$3\pi/16$ \quad  & $2.73 \pm 0.08$ \quad & $1.87\pm0.03$ \quad & $0.970 \pm 0.002$ \\
$\pi/4$ \quad & $2.65 \pm 0.06$ \quad & $1.84\pm0.01$ \quad & $0.970 \pm 0.002$ \\
$3\pi/8$ \quad & $2.85 \pm 0.10$ \quad & $1.78\pm0.06$ \quad & $0.982 \pm 0.002$ \\
\hline\hline
\end{tabular}
\end{center}
\end{table}

\subsection{Case I: Randomness on the links only}

We found two unstable fixed points on the boundaries of
parameter space~(\ref{eq: parameter space of network model})
and the expected phase diagram was shown in 
Fig.~\ref{fig: predicted phase diagram}.
Boundary $x=\infty$ realizes an insulating phase.
Boundary $\theta=0$ realizes the plateau transition 
at $x^{\ }_{\mathrm{cc}}=\ln(1+\sqrt{2})$
between two Hall insulating phases in the integer quantum Hall effect.
Boundaries $x=0$ and $\theta=\pi/2$ realize the unitary universality
class with its insulating phase terminating at the unstable metallic 
point $(x,\theta)=(0,\pi/2)$.
Any critical point close to the last three boundaries is difficult to identify
numerically as characteristic crossover length scales between different
universality classes become very large.
A related difficulty comes about from the fact that the
characteristic disorder strength can remain stronger than the characteristic
strength of the spin-rotation symmetry breaking away from the boundary 
$\theta=0$ of parameter space.\cite{Ando89}
For this reason, we use the scaling ansatz~(\ref{eq: scaling ansatz lambda'})
to search for the phase boundaries in the interior of
parameter space~(\ref{eq: parameter space of network model}).

\begin{figure}
\begin{center}
\includegraphics[width=8cm]{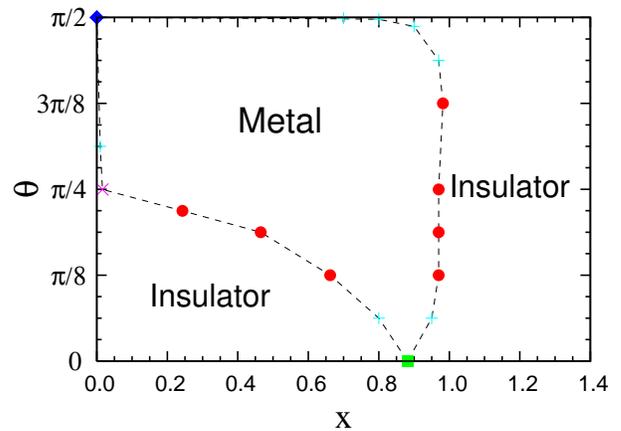}
\end{center}
\caption{
(Color online:)
Phase diagram for the network model 
in the parameter space~(\ref{eq: parameter space of network model}).
The location of a critical point 
denoted by a filled (red) circle follows from 
the scaling ansatz~(\ref{eq: scaling ansatz lambda'}).
That denoted by a cross is a rougher estimate due to large 
symmetry crossover effects.
The critical points denoted by a filled (green) square and 
a filled (blue) rhombus 
correspond to the critical points of the Chalker-Coddington
network model and the unstable metallic fixed point
from the unitary universality class, respectively.
Dashed lines are guide to the eye.
        } 
\label{fig:Phase_Diagram}
\end{figure}

For illustration, we present in 
Fig.\ \ref{fig:Lambda_theta=3pi/16}(a) 
the $x$ dependence of the normalized localization length $\Lambda$ 
for the fixed values of $\theta=3\pi/16$ and $M=4,8,16,32$. 
It is seen that $\Lambda$ increases with increasing $M$
for $x$ between $0.5$ and $0.9$.
For fixed $0.5<x<0.9$,
this is either the signature for an extended state or that for a localized
state whose localization length is larger than the maximal width of the
quasi-one dimensional network model.
Conversely, for $x$ smaller than $0.5$ or larger than $0.9$,
$\Lambda$ decreases with increasing $M$,
i.e., this is the signature of a localized state.
There appears to be two values of $x$, that we denote with
$x^{\ }_{s}<x^{\ }_{l}$, for which $\Lambda$ does not depend on 
$M=4,8,16,32$, and hence are good candidates for 
a pair of critical points separating a metallic from an insulating phase.
The inset of 
Fig.\ \ref{fig:Lambda_theta=3pi/16}(b) 
[Fig.\ \ref{fig:Lambda_theta=3pi/16}(c)]
magnifies the dependence of $\Lambda$ on $M=4,8,16,32,64$
close to $x^{\ }_{s}$ ($x^{\ }_{l})$.
On this scale $x^{\ }_{s}$ remains well-defined but not $x^{\ }_{l}$.
We attribute the absence of a single crossing point $x^{\ }_{l}$
in the inset of Fig.\ \ref{fig:Lambda_theta=3pi/16}(c)
to a large
contribution from an irrelevant scaling variable.
This hypothesis is verified
in Figs.\ \ref{fig:Lambda_theta=3pi/16}(b) 
and \ref{fig:Lambda_theta=3pi/16}(c) 
where the single-parameter dependence of $\Lambda'$
on the scaling variable 
$M^{1/\nu}|x-x^{\ }_{s}|$
and
$M^{1/\nu}|x-x^{\ }_{l}|$,
respectively,
is demonstrated (we found the value $y\approx-1$
for the largest irrelevant scaling exponent).
The values of
$\nu$, $\Lambda^{\ }_{c}$, $x^{\ }_{s}$, and $x^{\ }_{l}$
obtained from the scaling ansatz~(\ref{eq: scaling ansatz lambda'})
for different values of $\theta$ can be found in
Tables~\ref{tab:xs}
and~\ref{tab:xl}.
The values that we obtain for
$\nu$ and $\Lambda^{\ }_c$
are consistent with those for the standard two-dimensional symplectic 
universality class.%
\cite{Merkt98,Asada02} 
Our numerical map of the
phase boundaries separating the metallic from the insulating phase
in the parameter space~(\ref{eq: parameter space of network model})
is shown in
Fig.~\ref{fig:Phase_Diagram}.
The shape of the metallic region in Fig.~\ref{fig:Phase_Diagram}
is controlled by the crossover from the unitary 
to the symplectic symmetry class.

\begin{figure}[t]
\begin{center}
\includegraphics[width=8cm]{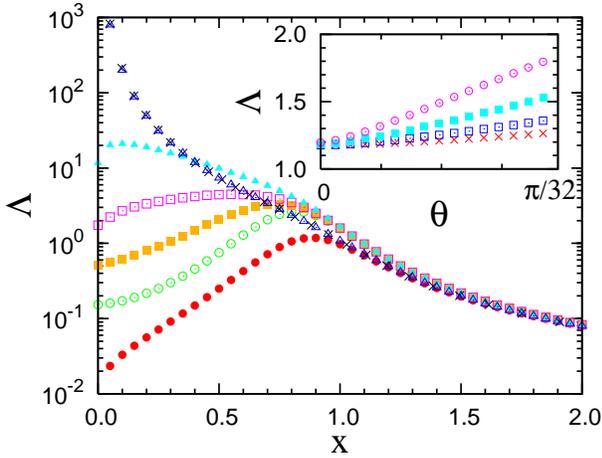}
\end{center}
\caption{
(Color online:)
Logarithm of the normalized localization length $\Lambda$
as a function of $x$ at fixed $M=16$ for
$\theta=0$ (filled circles), 
$\pi/8$ (open circles), 
$3\pi/16$ (filled squares), 
$\pi/4$ (open squares), 
$3\pi/8$ (filled triangles), 
and
$\pi/2$ (open triangles).
The duality relation~(\ref{eq: duality between x=0 and theta=piover2})
is verified by plotting the dependence
of $\ln\Lambda$ on $\theta$ when $x=0$ and $M=16$ (crosses)
as a function of $x=\mathrm{arctanh}(\cos\theta)$.
Inset: Dependence of $\Lambda$ on $\theta$ for 
$x=\ln(1+\sqrt{2})$ 
and 
$M=4$ (crosses),
$M=8$ (open squares), 
$M=16$ (filled squares), 
and
$M=32$ (open circles). 
        } 
\label{fig:Lambda_ThetaDep}
\end{figure}

The dependence of the normalized localization length $\Lambda$
on $\theta$ in the insulating regimes
$x<\ln(1+\sqrt{2})$ 
and
$x>\ln(1+\sqrt{2})$ 
are different as is shown in Fig.~\ref{fig:Lambda_ThetaDep}.
In the small $\theta$ insulating regime
$x<\ln(1+\sqrt{2})$
of Fig.~\ref{fig:Phase_Diagram},
$\Lambda$ is an increasing function of $\theta$
for fixed $x$ and $M$, as is expected from the proximity of
a phase boundary to a metallic phase.
In the insulating regime
$x>1>\ln(1+\sqrt{2})$
of Fig.~\ref{fig:Phase_Diagram},
$\Lambda$ depends weakly on $\theta$
for fixed $x$ and $M$, as is expected from a strongly localized regime.
When $x$ is held fixed at the Chalker-Coddington critical point
$x^{\ }_{\mathrm{cc}}=\ln(1+\sqrt{2})$.
$\Lambda$ is an increasing function of $\theta$ at fixed $M$
while $\Lambda$ is an increasing function of $M$ at fixed $\theta$.
This is the expected behavior assuming that any finite
$\theta$ drives the critical point
$(x,\theta)=\big(\ln(1+\sqrt{2}),0\big)$
into a metallic phase.
The duality relation~(\ref{eq: duality between x=0 and theta=piover2})
is also verified numerically in Fig.~\ref{fig:Lambda_ThetaDep}.

\subsection{Case II: Randomness on the links and nodes}

\begin{figure}[!]
\begin{center}
\includegraphics[width=8cm]{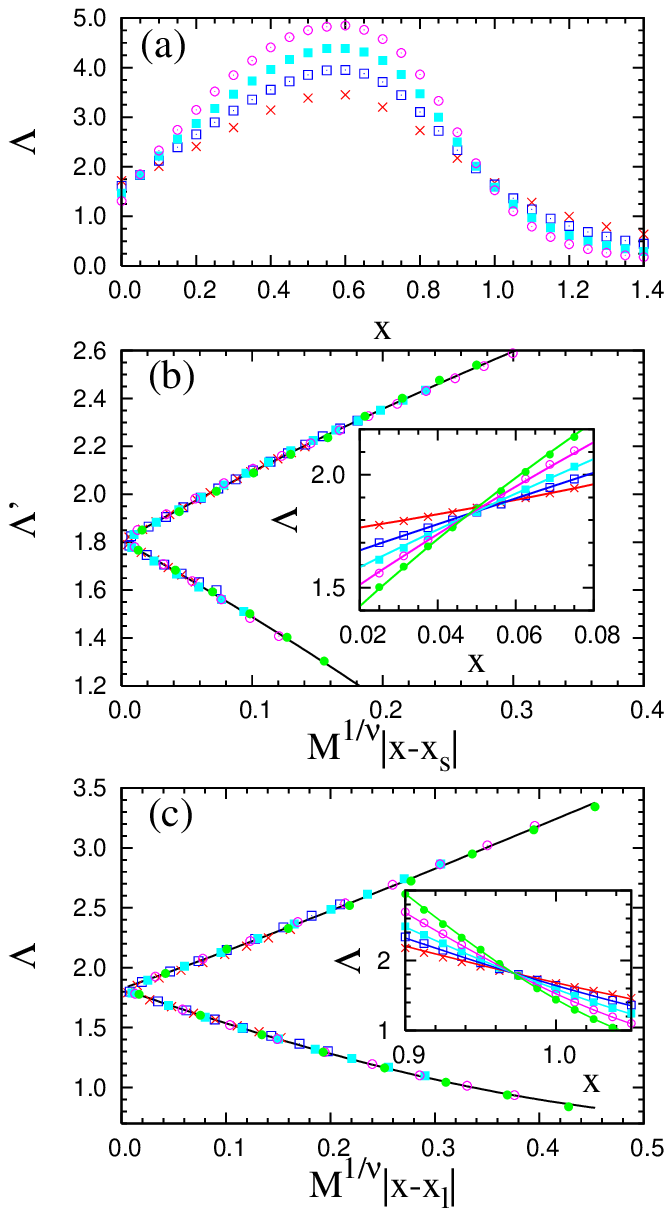}
\end{center}
\caption{
(Color online:)
(a) Normalized localization length $\Lambda$
as a function of $x$ with a random $\theta$
for the widths 
$M=4$ (times), 
$M=8$ (open squares), 
$M=16$ (filled squares), 
$M=32$ (open circles), 
and $64$ (filled circles)
of the quasi-one dimensional network.
Error bars are much smaller than symbol sizes.
There are two values of $x$ 
($x^{\ }_{s}\approx 0.05$ 
and 
$x^{\ }_{l}\approx  0.97$)
for which $\Lambda$ does not appear to depend on $M$.
(b) A fit of the data shown in (a)
with the help of the one-parameter 
scaling ansatz~(\ref{eq: scaling ansatz lambda'})
(whereby $f^{(\theta)}_{q,q}\to f^{\ }_{q,q}$
for all $p,q\in\mathbb{N}$) 
when $x$ is close to $x^{\ }_{s}$.
Inset: A blow up of (a) in the vicinity of $x^{\ }_{s}$. 
(c) A fit of the data shown in (a)
with the help of the one-parameter 
scaling ansatz~(\ref{eq: scaling ansatz lambda})
(whereby $f^{(\theta)}_{0,q}\to f^{\ }_{0,q}$
and $f^{(\theta)}_{1,q}=0$ for all $q\in\mathbb{N}$)
when $x$ is close to $x^{\ }_{l}$.
Inset: A blow up of (a) in the vicinity of $x^{\ }_{l}$. 
        }
\label{fig:Lambda_random}
\end{figure}

Following Asada et al.\ in Ref.~\onlinecite{Asada02},
we expect that corrections due to irrelevant scaling
variables should be reduced
by choosing $\theta$ to be independently distributed
between $0$ and $\pi/2$
with the probability $\sin(2\theta)$
for all nodes of the network.
As is illustrated with Fig.~\ref{fig:Lambda_random}, 
a metallic phase exists when $0.05 < x < 1.0$.
There are two quantum critical points $x^{\ }_{s}<x^{\ }_{l}$
separating the metallic phase $x^{\ }_{s}<x<x^{\ }_{l}$
from the insulating phase. The scaling analysis must account 
for an irrelevant scaling variable
with $y\approx-1$ in the vicinity of  $x^{\ }_{s}$.
In the vicinity of  $x^{\ }_{l}$ a single-parameter scaling analysis
suffices. Both scaling analysis, summarized in
Table~\ref{tab:random},
imply that the critical points 
$x^{\ }_{s}\approx0.05$ and $x^{\ }_{l}\approx0.97$
belong to the standard symplectic universality class.

\begin{table}[b]
\begin{center}
\caption{
Critical exponent, normalized localization length, and node parameter
when $\theta$ is distributed between $0$ and $\pi/2$.
        }
\label{tab:random}
\begin{tabular}{c c c c}
\hline\hline
 \quad & $\nu$ \quad & $\Lambda_c$ \quad & $x_c$ \\
\hline
$x_s$ \quad & $2.74 \pm 0.12$ \quad  & $1.81\pm0.01$ \quad & $0.047 \pm 0.001$ \\
$x_l$ \quad  & $2.68 \pm 0.06$ \quad & $1.82\pm0.01$ \quad & $0.971 \pm 0.001$ \\
\hline\hline
\end{tabular}
\end{center}
\end{table}

\section{Summary}
\label{sec: Summary}

We have constructed and studied a
two-dimensional spin-filtered chiral network model for the 
$\mathbb{Z}^{\ }_2$ quantum spin-Hall effect.
Disorder has been implemented in two distinct ways.
The quantum phase transitions between the
insulating and metallic state are found
to be characterized by the scaling 
exponent $\nu\approx2.7$ for the diverging localization length.
This value is consistent with that found in previous numerical
studies of the two-dimensional metal to insulator transition
in the symplectic universality class.\cite{Merkt98,Asada02}
We did not find the value $\nu\approx1.6$, 
recently observed by Onoda et al.\ in Ref.\ \onlinecite{Onoda06},
that was interpreted as the signature of a new
universality class at the transition between the
$\mathbb{Z}^{\ }_2$ quantum spin-Hall insulating and the metallic state.

It is important to remember the similarities and differences between
our network model and the lattice model studied in
Ref.\ \onlinecite{Onoda06}.
Common to the two models is that, in the absence of disorder,
they support a host $\mathbb{Z}^{\ }_{2}$ quantum spin-Hall state
(a host $\mathbb{Z}^{\ }_2$ insulator for brevity)
whereby an odd number of Kramers doublet
edge states cause an accumulation of
spin at the edges under an applied electric field.
The crucial difference lies in the spatial correlation of the
disorder potential added to the host $\mathbb{Z}^{\ }_2$ insulator.
On the one hand, in Ref.\ \onlinecite{Onoda06} disorder is introduced
as a random on-site potential that has no spatial correlation.
On the other hand, our network model is obtained by perturbing the
host $\mathbb{Z}^{\ }_2$ insulator with a spatially smooth disorder potential
that breaks the host $\mathbb{Z}^{\ }_2$ insulator
into droplets of $\mathbb{Z}^{\ }_2$ insulators.
The network model can thus be viewed as a coarse-grained effective model for
$\mathbb{Z}^{\ }_2$ insulating droplets 
that are weakly coupled through quantum tunneling.

The intrinsic symmetry (time reversal)
respected by the statistical ensemble of random 
Hamiltonians or scattering matrices 
is not changed by the range of the spatial correlation 
of the disorder. The hypothesis of universality 
would then suggest that the same critical scaling should be observed
at the localization-delocalization transitions
in the lattice model of Ref.\ \onlinecite{Onoda06} and in our
network model. The apparent violation of the universality 
by the two numerical results can be reconciled if one assumes
that there is a long crossover length scale beyond which
microscopic differences between the two models become irrelevant.
Corrections from irrelevant scaling variables may strongly depend
on the range of the disorder potential, 
as in the case of the plateau transition in the
second Landau level,\cite{Huckestein94}
and it could well be that the system sizes studied in 
Ref.\ \onlinecite{Onoda06}
were not large enough.
Verification of this scenario is left for future work.

The fact that our network model is built out of two Chalker-Coddington
network models coupled in a time-reversal invariant way has important
consequences.
Criticality in each Chalker-Coddington network model can be encoded 
by the field theory of a single (two-components) Dirac fermion coupled to
a random vector potential, a random mass, and a random scalar potential.%
\cite{Ho96}
It then follows by continuity (for small enough $\theta$)
that the two lines of critical points
emerging from the Chalker-Coddington critical point
in Fig.\ \ref{fig:Phase_Diagram}
can be encoded by a field theory for two flavors
of Dirac fermions. It is their coupling by disorder
that prevents the emergence of a time-reversal symmetric and
topologically driven quantum critical behavior.\cite{Ostrovsky07,Ryu07}

\section*{Acknowledgments}

We would like to thank Y.\ Avishai for useful discussions.
H.O., A.F., and C.M.\ acknowledge hospitality of the Kavli Institute for 
Theoretical Physics at Santa Barbara, where this work was initiated.
This work was supported by the
Next Generation Super Computing Project, Nanoscience Program, MEXT,
Japan and by the National Science Foundation
under Grant No.\ PHY99-07949.
Numerical calculations have
been mainly performed on the RIKEN Super Combined Cluster System.

\end{document}